\documentclass[onecolumn,trackchanges]{aastex701}

\usepackage{xcolor}  % for color support
\usepackage{xspace}
\usepackage{amsmath} 

\newcommand{\tess}{\textit{TESS}\xspace}
\newcommand{\tti}{\textit{TTI}\xspace}
\usepackage[utf8]{inputenc}
\usepackage{booktabs}

\begin{document}

\title{A Deep Precursor–Dip–Main Superoutburst Sequence in VW Hydri Observed with \tess: High-Cadence Constraints on the Thermal–Tidal Instability Model}

\author[orcid=0009-0001-7046-0446,sname='Ogunwale']{Ogunwale Bisi Bernard}
\affiliation{Department of Physics, Ariel University, Ariel 40700, Israel}
\affiliation{Kavli Institute for Astrophysics and Space Research, Massachusetts Institute of Technology, Cambridge, MA 02139, USA}
\email[show]{bisi.ogunwale@msmail.ariel.ac.il}  

\author[orcid=0000-0003-0155-2539, sname='Michael']{Michael Shara} 
\affiliation{Department of Astrophysics, American Museum of Natural History, New York, NY 10024, USA}
\email{mshara@amnh.org}

\author[orcid=0000-0002-1836-3120 , sname='Shporer']{Avi Shporer}
\affiliation{Kavli Institute for Astrophysics and Space Research, Massachusetts Institute of Technology, Cambridge, MA 02139, USA}
\email{shporer@mit.edu}

\author[orcid=0000-0002-7349-1109, sname='Guetta']{Dafne Guetta} 
\affiliation{Department of Physics, Ariel University, Ariel 40700, Israel}
\affiliation{Astrophysics, Geophysics, and Space Science Research Center, Ariel University, Ariel 40700, Israel}
\email{dafneg@ariel.ac.il}

\author[0000-0003-3757-1440,sname='Tal-Or']{Lev Tal-Or}
\affiliation{Department of Physics, Ariel University, Ariel 40700, Israel}
\affiliation{Astrophysics, Geophysics, and Space Science Research Center, Ariel University, Ariel 40700, Israel}
\email[show]{levtalor@ariel.ac.il}

\begin{abstract}
We present 120~s cadence \tess\ observations of three superoutbursts of the SU~UMa-type dwarf nova VW~Hydri. Two events (SO2 in Sectors~87+88 and SO3 in Sector~93) exhibit a pronounced, temporally pronounced precursor--dip followed by a rapid rise into the main superoutburst plateau. This morphology, previously seen in \textit{Kepler} light curves of V1504~Cyg and V344~Lyr, is a key prediction of the thermal--tidal instability (\tti) model when a normal (precursor) outburst expands the disk only marginally beyond the 3:1 resonance radius, allowing the tidal instability to grow slowly and produce a deep dip approaching quiescence before rapid amplification drives the main superoutburst. A sliding-window time--frequency analysis reveals superhump power already during the decline and near minimum light, with a smooth period evolution across the dip and stabilization after the system returns to the hot state, consistent with the growth and saturation of disk eccentricity at the 3:1 resonance. From the stabilized Stage~A superhump periods, we infer a representative mass ratio $q = 0.131 \pm 0.002$. Combined with either a typical SU~UMa white-dwarf mass prior or the semi-empirical donor sequence at an orbital period of 107~min, the implied component masses are $M_1 \simeq 0.6$--$1.0\,M_\odot$ and $M_2 \simeq 0.08$--$0.14\,M_\odot$, ruling out a brown-dwarf donor and establishing VW~Hyi as a benchmark system for testing tidal-instability models in low-$q$ dwarf novae.
\end{abstract}

%% You can use the \uat command to link your UAT concepts back to its source.

\section{Introduction} \label{sec:intro}

Dwarf novae (DNe) are a subclass of cataclysmic variable stars (CVs)—compact binaries in which a white dwarf accretes material from a Roche-lobe–filling late-type donor via an accretion disk. They exhibit recurrent optical outbursts driven by thermal instabilities in the disk \citep{Warner_1995}. Among these systems, the SU~UMa-type DNe occupy a key region below the 2–3\,h orbital-period gap, and display two distinct kinds of outbursts: short normal outbursts lasting a few days, and longer, brighter superoutbursts accompanied by photometric modulations known as superhumps \citep{Osaki_1989, Kato_2010, Osaki_Kato_2014}. Several normal outbursts typically occur between successive superoutbursts, defining the supercycle length. These systems provide an important observational testbed for accretion-disk instability theory and tidal interaction models in close binaries.

The leading theoretical framework for SU~UMa outbursts is the thermal–tidal instability (\tti) model \citep{Osaki_1989, Kato_2013}. In this picture, normal outbursts arise from thermal instabilities in the disk that episodically increase mass accretion onto the white dwarf. If, during one of these events, the disk expands sufficiently to reach the 3:1 orbital resonance radius \citep{Whitehurst_1988}, a tidal instability is triggered. This drives the disk into an eccentric, precessing state, enhancing tidal dissipation and powering a prolonged superoutburst. The eccentric disk produces superhumps, whose period reflects the beat between the orbital frequency and the slow apsidal precession of the outer disk \citep{Vogt_1982, Warner_1995}. Long-term alternation between Type~L (long) and Type~S (short) supercycles may reflect variations in the mass-transfer rate from the secondary star, potentially driven by stellar magnetic activity \citep{Warner_1995, Ichikawa1994}.

Superhump evolution during the superoutburst plateau is commonly divided into two stages. \textit{Stage~A} superhumps consist of low-amplitude modulations that grow rapidly during the rise or early plateau phase. Their longer periods are interpreted as tracing the dynamical precession rate at the 3:1 resonance, providing a direct diagnostic of the binary mass ratio \citep{Kato_2010, Osaki_Kato_2013}. As the disk settles into a fully developed eccentric configuration, \textit{Stage~B} superhumps dominate, exhibiting slightly shorter and more stable periods \citep{Patterson_2005, Kato_2010, Osaki_Kato_2014}.
% ====================== TESS GALLERY: YEAR 1 ======================
\begin{figure*}[t!]
    \centering
    \includegraphics[width=0.95\linewidth]{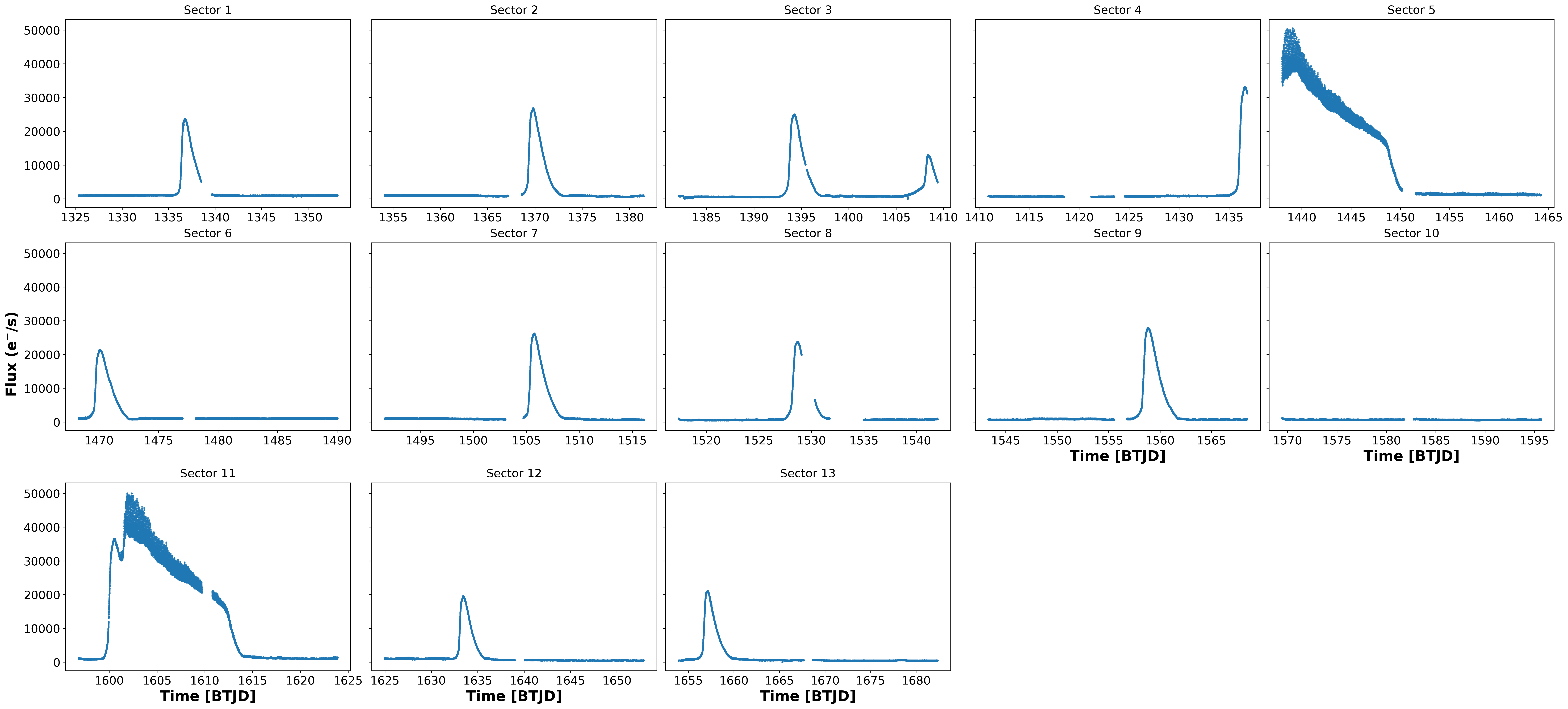}
    \caption{
        \textbf{TESS Primary Mission (Year 1) Light curve.}
        \tess\ SPOC Simple Aperture Photometry (SAP) raw flux light curves of VW~Hyi
        across Sectors 1--13. Each panel shows the raw flux as a function of time
        in TESS BJD (BTJD = BJD $-$ 2457000).
    }
    \label{fig:vwhyi_sap_year1}
\end{figure*}

% ====================== TESS GALLERY: YEAR 3 (CYCLE 3) ======================
\begin{figure*}[t!]
    \centering
    \includegraphics[width=0.95\linewidth]{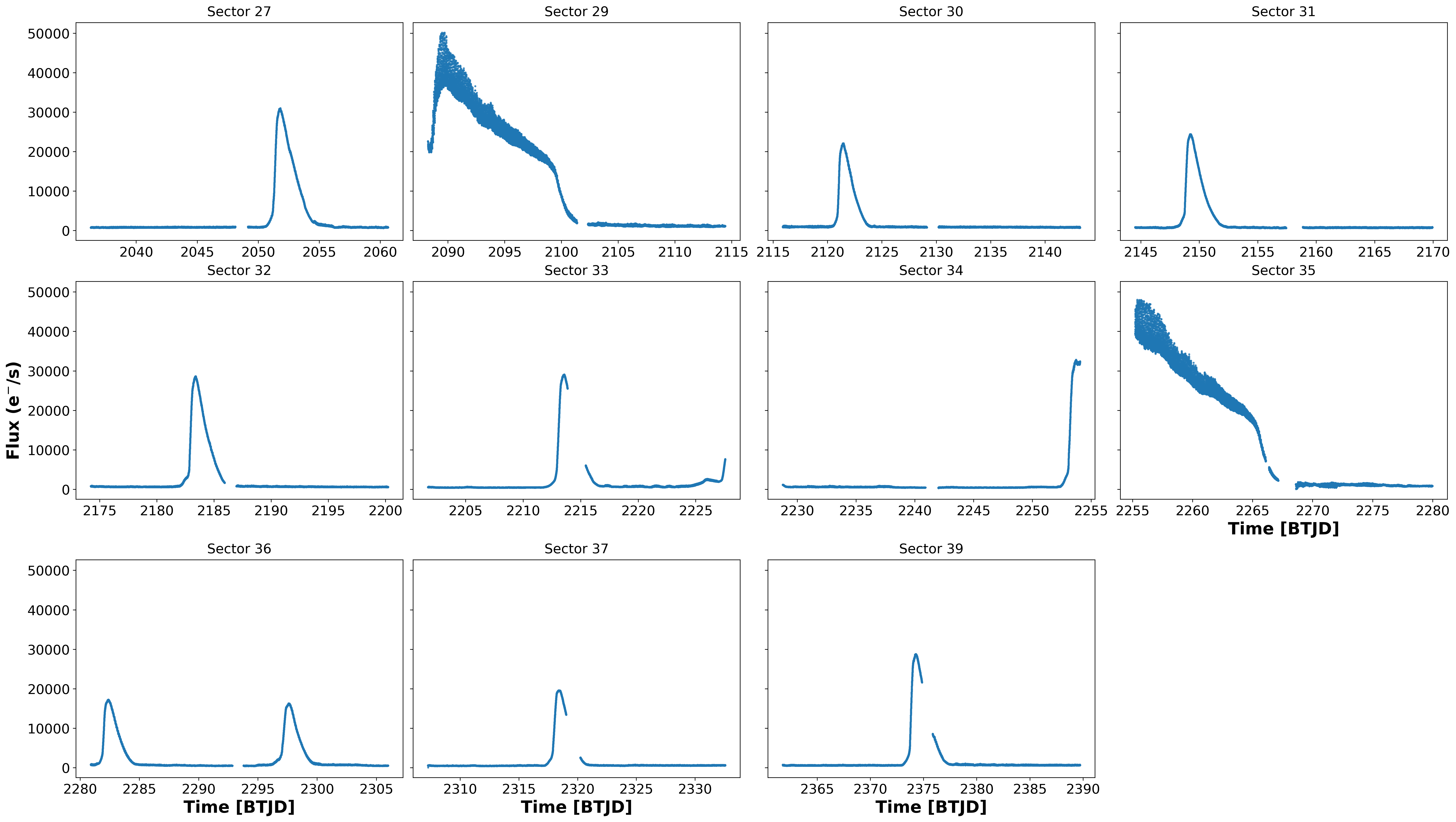}
    \caption{
        \textbf{TESS First Extended Mission (Year 3, Cycle 3) light curve.}
        \tess\ SPOC SAP light curves of VW~Hyi covering Sectors 27 and 29--37.
    }
    \label{fig:vwhyi_sap_year2}
\end{figure*}

% ====================== TESS GALLERY: YEAR 5 (CYCLE 5) ======================
\begin{figure*}[t!]
    \centering
    \includegraphics[width=0.95\linewidth]{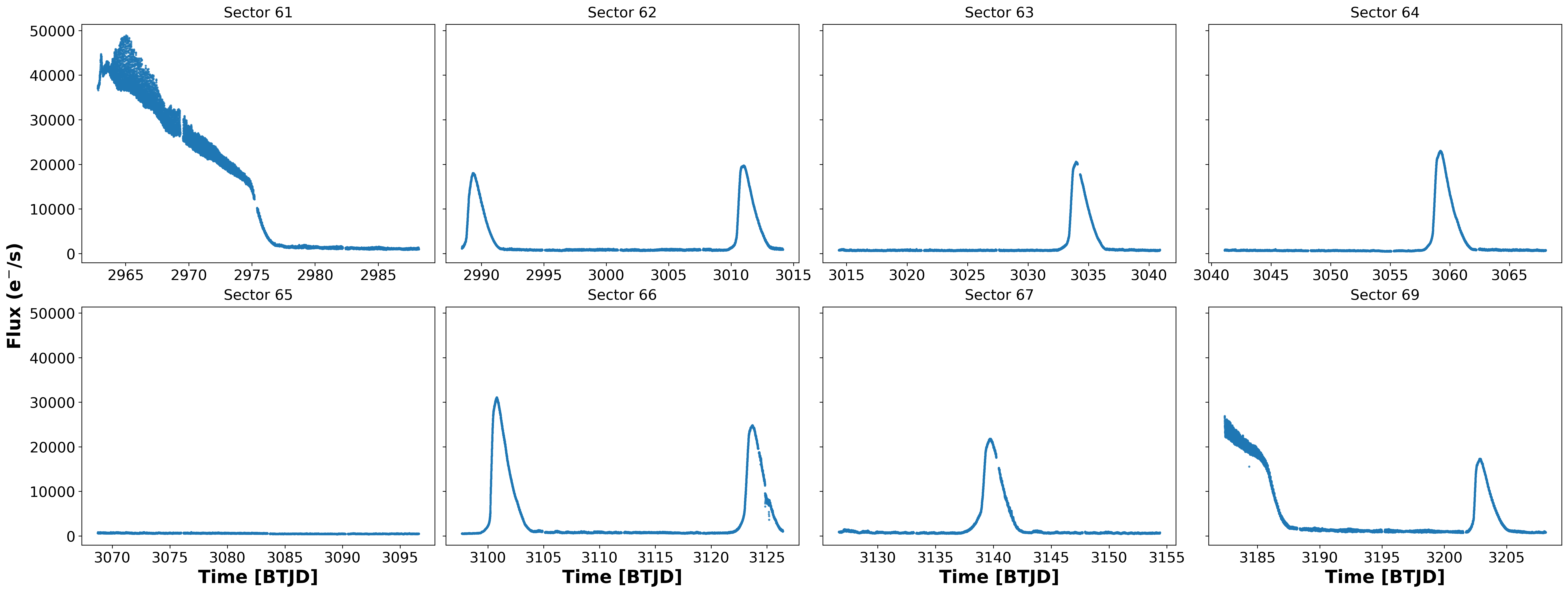}
    \caption{
        \textbf{TESS Second Extended Mission (Year 5, Cycle 5) light curve.}
        \tess\ SPOC SAP light curves of VW~Hyi across Sectors 61--69.
    }
    \label{fig:vwhyi_sap_year3}
\end{figure*}

% ====================== TESS GALLERY: YEAR 7 (CYCLE 7) ======================
\begin{figure*}[t!]
    \centering
    \includegraphics[width=0.95\linewidth]{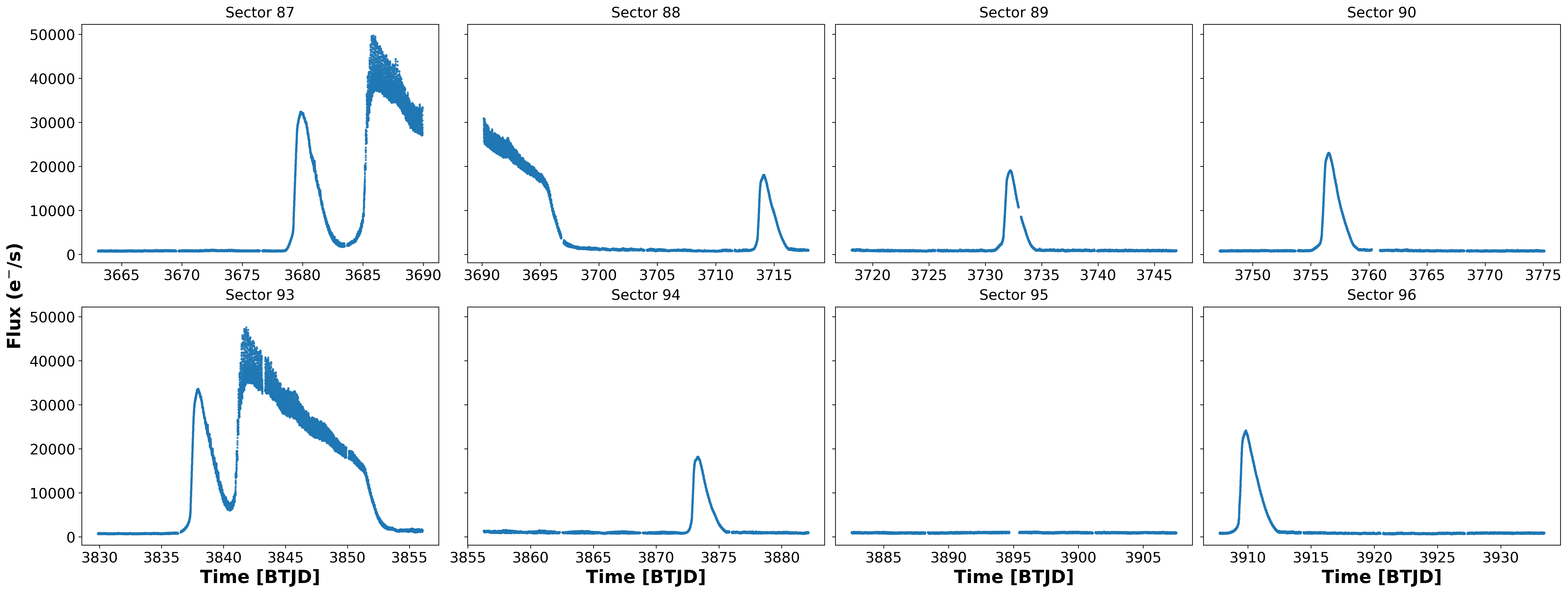}
    \caption{
        \textbf{TESS Third Extended Mission (Year 7, Cycle 7) light curve.}
        \tess\ SPOC SAP light curves of VW~Hyi across Sectors 87--96.
    }
    \label{fig:vwhyi_sap_year4}
\end{figure*}

VW~Hyi is a prototypical and extensively studied SU~UMa-type dwarf nova. Its orbital period has been independently measured through both photometric and spectroscopic methods, with remarkable consistency over nearly five decades (Table~\ref{tab:vw_hyi_periods}). The first reliable determination was obtained by \citet{Vogt_1974}, who identified a stable photometric hump at $0.07427111$\,d (approximately 107\,min). This value was subsequently confirmed by radial-velocity measurements \citep{Schoembs_1981}. Doppler tomography by \citet{Smith_2006} established the phase of inferior conjunction of the donor, revealing a small offset relative to the photometric ephemeris. Most recently, \citet{Saez_Vogt_2023} refined the period using Gaia and coordinated ground-based photometry, further reinforcing the canonical value. The close agreement among these determinations highlights the long-term dynamical stability of the system.

VW~Hyi resides in the southern continuous viewing zone (CVZ) of the \tess mission ($\alpha = 62.2983^{\circ}$, $\delta = -71.2949^{\circ}$), providing an exceptional opportunity for high-cadence, long-baseline monitoring. Its light curves reveal both normal and superoutbursts, and decades of observations have shown alternating Type~L supercycles ($\sim$250–300\,d with many normal outbursts) and Type~S supercycles ($\sim$180–190\,d with fewer normal outbursts) \citep{Warner_1995, Bateson_1977, Otulakowska2013}. This pronounced and well-characterized variability makes VW~Hyi a benchmark system for testing models of dwarf-nova outbursts and for developing tidal instability models.

Figures~\ref{fig:vwhyi_sap_year1}–\ref{fig:vwhyi_sap_year4} present the \tess Simple Aperture Photometry (SAP) light curves of VW~Hyi across the mission’s primary and extended phases. The continuous monitoring at 120-second cadence provides the highest time-resolution space-based observations of this system to date.

\begin{table}[ht]
\centering
\caption{Published determinations of the orbital period of VW~Hyi.}
\begin{tabular}{llll}
\hline
\textbf{Reference} & \textbf{Period (days)} & \textbf{Method} & \textbf{Notes} \\
\hline
\citet{Vogt_1974} & 0.07427111 & Photometric & Quiescent hump; epoch from 1973 \\
\citet{Schoembs_1981} & $\sim$0.074271 & Spectroscopic & Radial velocities consistent with photometry \\
\citet{Smith_2006} & $\sim$0.07427 & Spectroscopic & Doppler tomography; phase offset from hump \\
\citet{Saez_Vogt_2023} & 0.07427130(14) & Photometric & Gaia + ground-based; refined precision \\
\hline
\end{tabular}
\label{tab:vw_hyi_periods}
\end{table}

In this paper, we present a detailed analysis of VW~Hyi using \tess short-cadence observations. Section~\ref{sec:observation} describes the \tess data and reduction procedures. Section~\ref{sec:results} presents our measurements of superhump evolution, mass ratio, superoutburst morphology, and energetic properties. Section~\ref{sec:discussion} discusses the broader implications for the thermal–tidal instability model and comparisons with other dwarf novae. We summarize our conclusions in Section~\ref{sec:conclusion}.

\section{\tess Observations} \label{sec:observation}

The Transiting Exoplanet Survey Satellite \citep[\tess;][]{Ricker_2015} was launched in 2018 into a highly stable elliptical Earth orbit \citep{Gangestad_2013} with the primary goal of detecting transiting exoplanets around bright nearby stars. Its four wide-field 10\,cm cameras, each comprising four CCDs, provide continuous 24\textdegree\,$\times$\,96\textdegree\ sky coverage for $\sim$27\,days per pointing (a ``Sector''). Although designed for exoplanet discovery, the mission’s large field of view, high duty cycle, and multi-year baseline have made it a premier facility for time-domain astrophysics, enabling detailed studies of variable stars, compact binaries, and optical transients \citep[e.g.,][]{Wong_2021, Fausnaugh_2023, Pras_2022, Antoci_2019}.

\tess delivers two primary classes of photometric data: (1) short-cadence target pixel files (TPFs) for pre-selected targets, and (2) full-frame images (FFIs) with cadences of 1800\,s, 600\,s, or 200\,s, depending on the observing year.  
Short-cadence light curves are processed by the \tess Science Processing Operations Center (SPOC) pipeline \citep{Jenkins_2016}, while FFI-based light curves are produced by the Quick Look Pipeline \citep[QLP;][]{Huang_2020, Kunimoto_2021} and by independent efforts such as T16 \citep{Hartman_2025} and TGLC \citep{Han_2023}. All these products are publicly accessible through MAST.\footnote{\url{https://archive.stsci.edu/missions-and-data/tess}}

\citet{Ogunwale_2025} released a catalog of variable-source light curves extracted from the first two years of \tess\ FFIs, now hosted as a MAST High-Level Science Product.\footnote{\url{https://archive.stsci.edu/hlsp/tequila/}}   
This catalog served as our starting point for identifying systems exhibiting large-amplitude or rapid brightness changes indicative of transient or eruptive behavior. By scanning the catalog for such variability signatures, we isolated several promising candidates, including multiple dwarf novae. For each candidate, we then used \texttt{TESSPoint} \citep{Burke_2020} to determine its precise detector coordinates and to map its observational coverage across all available \tess sectors. 
This procedure revealed that VW~Hyi had extensive short-cadence observations spanning multiple mission years, making it an ideal case for a detailed investigation of superoutburst morphology and superhump evolution in high-cadence \tess\ data.

The SPOC pipeline provides two photometric time series: Simple Aperture Photometry (SAP) and Presearch Data Conditioning SAP (PDCSAP). PDCSAP applies corrections for instrumental systematics—such as pointing jitter, scattered light, and thermal trends—using cotrending basis vectors constructed from quiet stars on the same CCD channel \citep{Smith_2012, Stumpe_2012, Stumpe_2014}. Although PDCSAP is typically preferred for transit searches and short-period signals, its detrending can suppress or distort long-term variability, particularly in systems with large intrinsic amplitude changes \citep{Van_Cleve_2016}. Because preserving the full outburst morphology is essential for our analysis of VW~Hyi, we therefore adopt the uncorrected SAP fluxes.

In this work, we focus exclusively on the 120\,s short-cadence SAP light curves, which provide the highest temporal resolution available for VW~Hyi. We scale the light curves by the \texttt{FLFRCSAP} keyword provided on the header of each light curve file to account for the fractional flux count loss in the SAP photometry. The full set of \tess of the scaled SAP observations used in this study is shown in Figures~\ref{fig:vwhyi_sap_year1}–\ref{fig:vwhyi_sap_year4}.

\section{Results}\label{sec:results} 

\subsection{Superhump Detection and Superoutburst Characterization}
\label{sec:superhump_detection_and_superoutbursts}

To identify and characterize superhumps in the \textit{TESS} light curve of VW~Hyi, we performed a sliding-window Lomb--Scargle (LS) periodogram analysis across the full observational baseline. Local power spectra were computed in windows of fixed duration ($\Delta t = 1.5$~d), stepped by 0.25~d, allowing us to track the temporal evolution of
periodic signals during outburst episodes. Candidate superhump detections were selected by requiring (i) periods in the range $0.075$--$0.078$~d, (ii) a false-alarm probability below $10^{-3}$, and (iii) temporal coincidence with visually identified bright outbursts in the light curve. Individual detections were grouped using the DBSCAN clustering
algorithm \citep{Ester1996}, which associates temporally and spectrally coherent detections into distinct events. Using this approach, we identify eight superoutbursts across the \textit{TESS} baseline, occurring in Sectors~5, 11, 29, 35, 61, 69, 87+88, and~93. Among these, three events (Sectors~11, 87+88, and~93) are particularly well sampled,
with complete coverage of the precursor, plateau, and decay phases, as well as clearly resolved superhump evolution. For the time--frequency maps in Figures~\ref{fig:SO2_VW_Hyi_dynamic_LS} and \ref{fig:SO3_VW_Hyi_dynamic_LS}, we recompute the sliding-window LS spectra on the superoutburst intervals using parameters optimized for visualization; the specific choices are given in the figure captions.

\begin{figure*}
  \centering
  \includegraphics[width=17.5cm, height=12.5cm]{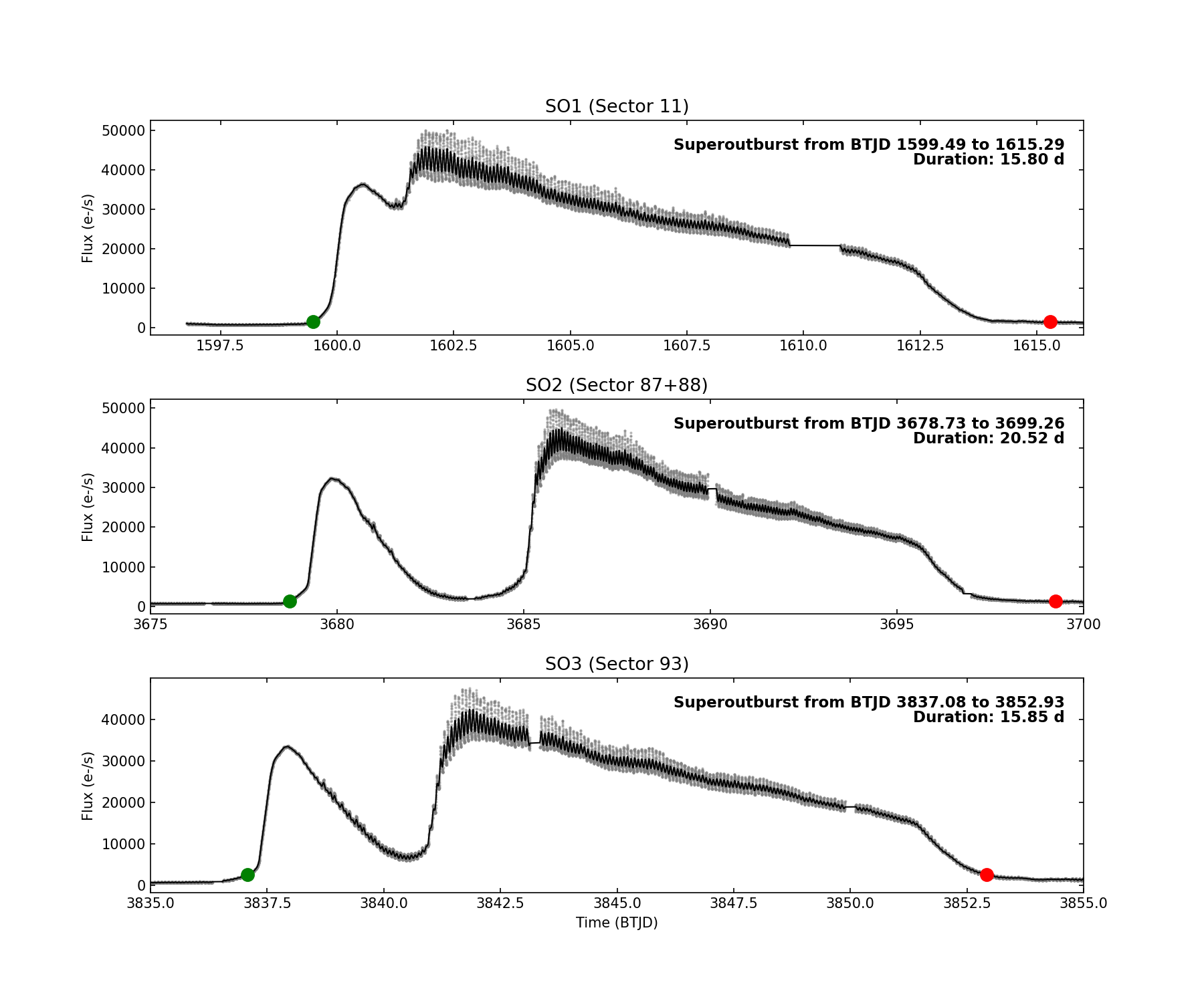}
  \caption{Three superoutbursts of VW~Hyi recorded by \textit{TESS}, showing the precursor, main plateau, and decay phases. Green and red markers indicate the start and end of the superoutburst episode.}
  \label{fig:superoutburst}
\end{figure*}

To measure superoutburst durations, we adopted a time-domain thresholding method intended to capture the full outburst episode, including the precursor (normal) outburst. For each event, the light curve was smoothed with a 1D Gaussian filter with kernel standard deviation $\sigma_{\rm G}=15$ samples (corresponding to $\sim$30~minutes at the 2-minute
\textit{TESS} cadence) to suppress short-timescale variability. A quiescent baseline flux level was estimated using sigma-clipped statistics of the smoothed flux distribution, and a brightness threshold was defined as
\[
F_{\mathrm{thresh}} = F_{\mathrm{base}} + 2.5\,\sigma_{\mathrm{base}},
\]
where $F_{\mathrm{base}}$ is the median baseline flux and $\sigma_{\mathrm{base}}$ is the sigma-clipped standard deviation. We then identified above-threshold intervals in the smoothed light curve and defined the superoutburst onset and termination as the first and last crossings of the dominant above-threshold interval encompassing the event. With
this choice, the precursor/normal outburst and the subsequent main superoutburst plateau are treated as a single contiguous episode (i.e., the superoutburst duration is measured from the onset of the precursor/normal outburst through the final decay back to quiescence). The three best-observed superoutbursts are shown in Figure~\ref{fig:superoutburst}, with the inferred start and end times marked. These events form the basis for our detailed analysis of superhump evolution and mass-ratio determination.

\subsection{Stage-A Superhump Periods and the Mass Ratio of VW~Hyi}
\label{sec:superhump_periods_mass_ratios}

To determine the binary mass ratio $q$ of VW~Hyi, we measured the Stage-A superhump periods during the three well-observed superoutbursts in Sectors~11, 87+88, and~93. We adopt an orbital period of $P_{\mathrm{orb}} = 0.074271$~d from spectroscopic measurements \citep{Schoembs_1981}. Stage-A superhumps correspond to the earliest phase of superhump development and reflect the nearly pure dynamical apsidal precession rate at the 3:1 resonance radius, with minimal contamination from pressure effects \citep[e.g.,][]{Osaki_Kato_2013}. Consequently, they provide the most reliable method for determining the mass ratio in SU~UMa-type dwarf novae.

\begin{figure*}
  \centering
  \includegraphics[width=12.5cm, height=10.0cm]{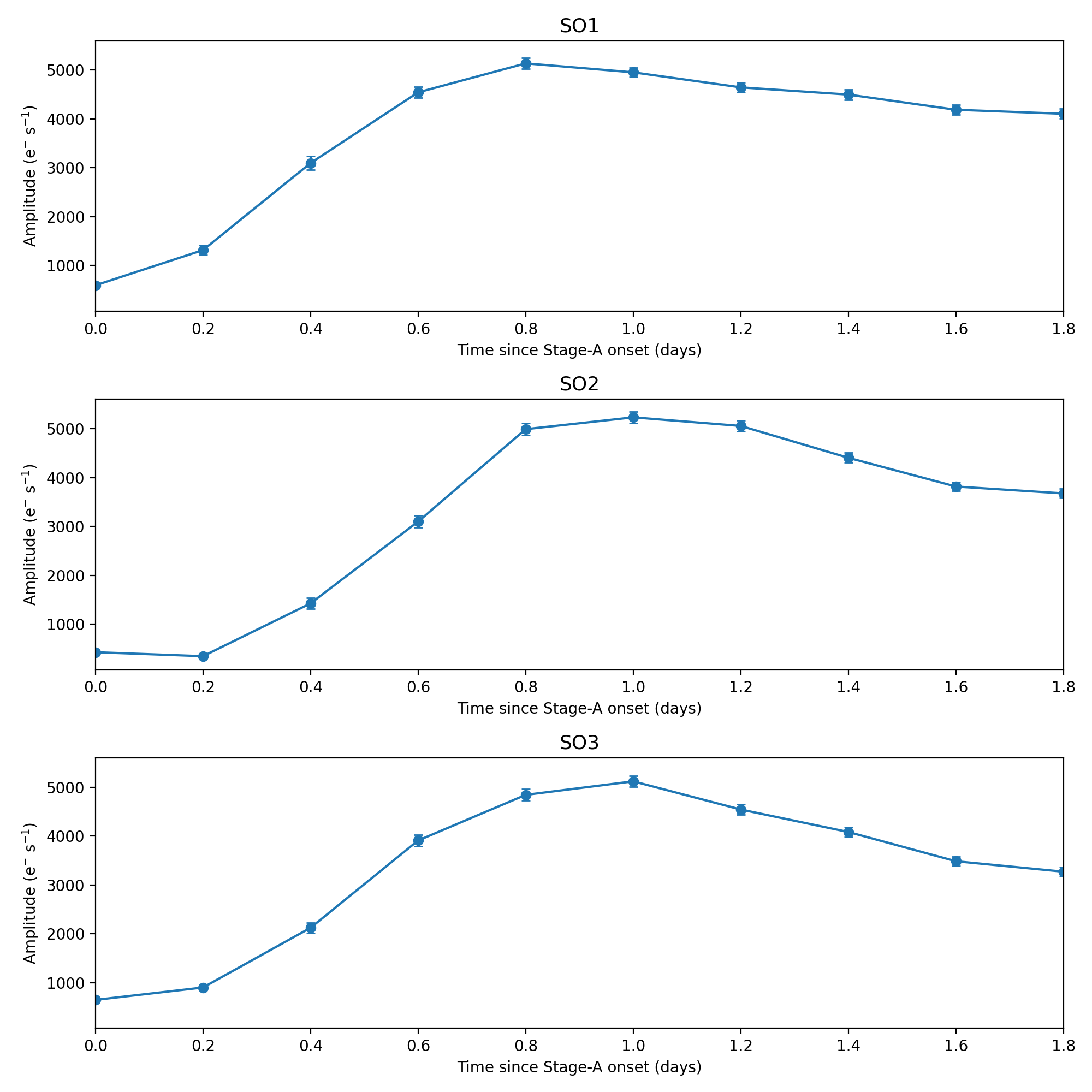}
  \caption{Temporal evolution of the superhump amplitude during the early stages of the three well-observed \textit{TESS} superoutbursts of VW~Hyi. In each panel, the light curve is divided into overlapping time windows, locally detrended, and fit with a sinusoid at the fixed Stage-A superhump period. The rapid growth phase identifies Stage~A, followed by saturation corresponding to Stage~B. Time is referenced to the onset of Stage~A in each event.}
  \label{fig:sh_amplitude_evolution}
\end{figure*}

For each superoutburst, we isolated the earliest interval in which the superhump signal first emerges and grows rapidly in amplitude. Within each window, we detrended the light curve by subtracting a Savitzky--Golay smooth (quadratic; $\sim0.4$~d window) and computed a Lomb--Scargle periodogram on the residuals over a narrow band around a prior estimate. The strongest peak prior to the plateau defines the Stage-A superhump period $P_{\mathrm{sh,A}}$. To validate the Stage-A identification, we tracked the superhump amplitude in overlapping windows by fitting a sinusoid at a fixed trial period; in all three events the amplitude rises monotonically for $\sim0.8$--$1.0$~d before saturating as Stage~B begins (Figure~\ref{fig:sh_amplitude_evolution}). We therefore restrict $P_{\mathrm{sh,A}}$ measurements to the rising portion of these amplitude curves.

To estimate realistic uncertainties on $P_{\mathrm{sh,A}}$, we adopt a window-sensitivity approach. Specifically, for each superoutburst, we measure the Stage-A period in two slightly offset but overlapping Stage-A windows and adopt the scatter between the two measurements as the primary uncertainty estimate. This approach captures the dominant source of uncertainty associated with the finite duration of the Stage-A phases of the events.

Using this method, we measure Stage-A superhump periods of $P_{\mathrm{sh,A}} = 0.077886 \pm 0.00012$~d (SO1), $0.077778 \pm 0.00013$~d (SO2), and
$0.077801 \pm 0.00006$~d (SO3).
To infer the mass ratio, we compute the fractional superhump excess in frequency,
\begin{equation}
\epsilon^{*} = 1 - \frac{P_{\mathrm{orb}}}{P_{\mathrm{sh,A}}},
\end{equation}
and apply the dynamical Stage-A calibration of \citet{Kato_Osaki_2013}, implemented using the polynomial approximation
\begin{equation}
q = -0.0016 + 2.60\,\epsilon^{*} + 3.33\,(\epsilon^{*})^2 + 79.0\,(\epsilon^{*})^3,
\end{equation}
which approximates the calibration to a maximum error of $\Delta q \simeq 4\times10^{-4}$
over $0.025 \le q \le 0.394$.
For our measured periods we obtain $\epsilon^{*}=0.04641$ (SO1), $0.04509$ (SO2), and $0.04537$ (SO3),
corresponding to mass ratios
$q_A = 0.1341 \pm 0.0047$, $0.1296 \pm 0.0054$, and $0.1306 \pm 0.0024$, respectively. The uncertainties are dominated by the
Stage-A period measurement rather than by the polynomial approximation itself.

The mass ratios derived independently from the three superoutbursts are mutually consistent and fall within the expected range for VW~Hyi. In particular, \citet{Kato_etal_2014} and \citet{Kato_2022} report $q = 0.126 \pm 0.005$ for VW~Hyi using the Stage-A superhump formalism based on ground-based observations. Our \textit{TESS}-based results, derived from continuous high-cadence space-based photometry, agree within uncertainties and provide an independent confirmation of these earlier measurements.

For comparison, we also list mass ratios inferred from Stage-B superhumps using the empirical period-excess relation of \citet{Patterson_2005}. As expected, the Stage-B estimates are systematically larger, reflecting the influence of pressure effects on the disk precession rate \citep[e.g.,][]{Lubow_1992}. We therefore adopt the Stage-A mass ratios as our preferred values. The final superhump periods and mass-ratio estimates are summarized in Table~\ref{tab:vwhyi_superhumps}.

\begin{table}
\centering
\caption{Superhump periods and inferred mass ratios for VW~Hyi from \tess.}
\label{tab:vwhyi_superhumps}

\begin{tabular}{l l c c c c}
\hline
Event & Sector(s) & Stage & $P_{\rm sh}$ (d) & Excess & $q$ \\
\hline
SO1 & 11    & A & $0.077886 \pm 0.00012$ & $0.04641$ & $0.1341 \pm 0.0047$ \\
SO1 & 11    & B & $0.07683 \pm 0.00026$  & $0.03440$ & $0.1533 \pm 0.0129$ \\
SO2 & 87+88 & A & $0.077778 \pm 0.00013$ & $0.04509$ & $0.1296 \pm 0.0054$ \\
SO2 & 87+88 & B & $0.07668 \pm 0.00015$  & $0.03245$ & $0.1460 \pm 0.0075$ \\
SO3 & 93    & A & $0.077801 \pm 0.00006$ & $0.04537$ & $0.1306 \pm 0.0024$ \\
SO3 & 93    & B & $0.07672 \pm 0.00025$  & $0.03296$ & $0.1479 \pm 0.0125$ \\
\hline
\end{tabular}

{\footnotesize \emph{Notes.} $P_{\rm orb}=0.074271$~d. For Stage~A we list the frequency excess
$\epsilon^{*}=1-P_{\rm orb}/P_{\rm sh,A}$ and compute $q$ using the dynamical calibration of \citet{Osaki_Kato_2013} (polynomial approximation). For Stage~B we list the period excess $\epsilon=(P_{\rm sh,B}-P_{\rm orb})/P_{\rm orb}$ and compute $q$ using the empirical relation of \citet{Patterson_2005}.}
\end{table}

\subsection{Disk Radius and Precession Period}
\label{sec:disk_radius_precession}

As described in Section~\ref{sec:superhump_periods_mass_ratios}, the positive-superhump signal evolves through the canonical Stage~A and Stage~B phases. To estimate the disk apsidal precession period, we therefore adopt the \emph{plateau} Stage~B superhump period, measured by restricting the LS analysis to the main superoutburst plateau ( Table~\ref{tab:vwhyi_superhumps}). A weighted average over the three events yields
\[
P_{\mathrm{sh,B}} = 0.076717 \pm 0.000113~\mathrm{d},
\]
where the uncertainty reflects the robust scatter among overlapping plateau subwindows and the event-to-event dispersion.

Using the spectroscopic orbital period $P_{\mathrm{orb}}=0.074271$~d \citep{Schoembs_1981}, the apsidal precession period implied by the superhump--orbital beat is
\[
P_{\mathrm{prec}} =
\left|
\frac{1}{P_{\mathrm{orb}}} - \frac{1}{P_{\mathrm{sh,B}}} \right|^{-1}
= 2.33 \pm 0.10~\mathrm{d}.
\]
A modulation on this timescale is consistent with the slow amplitude and phase variations often present in superhump light curves during the plateau phase. The location of the 3{:}1 eccentric Lindblad resonance in a Keplerian disk is
\[
\frac{r_{3{:}1}}{a} = 3^{-2/3}(1+q)^{-1/3},
\]
where $a$ is the binary separation and $q$ is the mass ratio \citep[e.g.,][]{Whitehurst_1988}. Substituting our preferred Stage~A mass ratio (representative value $q = 0.131 \pm 0.002$) gives
\[
\frac{r_{3{:}1}}{a} = 0.4614 \pm 0.0003 .
\]
This characteristic radius is consistent with the standard picture in which the accretion disk in VW~Hyi expands to (or slightly beyond) the 3{:}1 resonance radius during superoutburst, triggering the tidal instability responsible for positive-superhump formation and the associated apsidal precession of the eccentric disk \citep[e.g.,][]{Hirose_Osaki_1990}.

\begin{deluxetable*}{lccccc}
\tablecaption{\tess-band isotropic-equivalent energy estimates for the three VW~Hyi superoutbursts, with an illustrative bolometric-energy range.\label{tab:Etess}}
\tablecolumns{6}
\tablewidth{0pt}
\tablehead{
\colhead{Superoutburst} &
\colhead{Duration Interval} &
\colhead{Duration} &
\colhead{$\mathcal{F}_{\mathrm{TESS}}$} &
\colhead{$E_{\mathrm{TESS}}$} &
\colhead{$E_{\mathrm{bol}}$} \\
\colhead{} &
\colhead{(BTJD)} &
\colhead{(d)} &
\colhead{($10^{-3}\,\mathrm{erg\,cm^{-2}}$)} &
\colhead{($10^{38}\,\mathrm{erg}$)} &
\colhead{(BC$\!=34$--$320$; $10^{40}\,\mathrm{erg}$)}
}
\startdata
SO1 (Sector 11)      & 1599.48974 -- 1615.28585 & 15.78 & 1.147 & 3.960 & 1.346--12.67 \\
SO2 (Sectors 87+88)  & 3678.73367 -- 3699.25570 & 20.52 & 1.167 & 3.993 & 1.358--12.78 \\
SO3 (Sector 93)      & 3837.08217 -- 3852.92823 & 15.85 & 1.103 & 3.807 & 1.294--12.18 \\
\enddata
\tablecomments{
Energy estimates use a nominal \tess\ effective wavelength $\lambda_{\mathrm{eff}}=775.5$~nm. The detected energy is converted to  fluence via $\mathcal{F}_{\rm TESS}=E_{\rm det}/A_{\rm eff}$ adopting an on-axis effective area $A_{\rm eff}=70~\mathrm{cm^2}$. The isotropic-equivalent band energy is $E_{\mathrm{TESS}} = 4\pi d^2 \mathcal{F}_{\rm TESS}$ with $d=53.7$~pc. The bolometric range assumes $E_{\mathrm{bol}} \approx \mathrm{BC}\,E_{\mathrm{TESS}}$ with $\mathrm{BC}=34$--$320$ (corresponding to $T_{\rm eff}\approx 20{,}000$--$50{,}000$~K).
}
\end{deluxetable*}

\subsection{Outburst Duration and \tess-band Energetics}
\label{sec:energetics}

In this section, we quantify the \tess-band energetics by measuring the quiescent-subtracted fluence in the corrected \texttt{SAP\_FLUX} light curves and converting it to an isotropic-equivalent band energy under explicitly stated instrumental assumptions.

For each event, we define the superoutburst interval as described in Section~\ref{sec:superhump_detection_and_superoutbursts} and listed in Table~\ref{tab:Etess}. We estimate a quiescent baseline level, $F_{\rm base}$, from a 2-day window immediately preceding the outburst, adopting the median flux. If pre-outburst coverage is insufficient, we instead adopt the 10th percentile of the available flux distribution.

We define the quiescent-subtracted excess flux as
\[
F_{\rm exc}(t) = \max\!\left[F(t)-F_{\rm base},\,0\right],
\]
and compute the excess electron fluence by integrating $F_{\rm exc}(t)$ over the superoutburst window, yielding the total detected fluence, $\mathcal{N}_e$. This corresponds to the total number of detected electrons attributable to the outburst above quiescence within the \tess\ bandpass.

To place the fluence on an energy scale, we adopt a harmonic-mean effective wavelength for the \tess\ bandpass of $\lambda_{\rm eff}=775.5$~nm and compute the energy per photon as
\[
E_{\gamma}=\frac{hc}{\lambda_{\rm eff}}.
\]
We then define the detected \tess-band energy at the instrument as
\[
E_{\rm det} = \mathcal{N}_e\,E_{\gamma}.
\]

Next, we convert to an energy fluence at Earth via
\[
\mathcal{F}_{\mathrm{TESS}} = \frac{E_{\rm det}}{A_{\rm eff}},
\]
where $A_{\rm eff}$ is the on-axis effective collecting area of the \tess\ camera. Using the published \tess\ throughput curves and wavelength-dependent effective area \citep{Vanderspek_2018}, we adopt a nominal value of $A_{\rm eff}=70\,\mathrm{cm}^2$ at the peak of the bandpass. Finally, we compute the isotropic-equivalent \tess-band energy as
\begin{equation}
E_{\mathrm{TESS}} = 4\pi d^2\,\mathcal{F}_{\mathrm{TESS}},
\end{equation}
adopting $d=53.7$~pc from the \emph{Gaia} distance estimate.

Uncertainties on $\mathcal{N}_e$ are estimated using 200 Monte--Carlo realizations in which the in-window fluxes are perturbed by Gaussian noise drawn from \texttt{SAP\_FLUX\_ERR}, while holding $F_{\rm base}$ fixed. The resulting scatter in $\mathcal{N}_e$ is propagated linearly to $E_{\rm det}$, $\mathcal{F}_{\oplus,\mathrm{TESS}}$, and $E_{\mathrm{TESS,iso}}$. The results obtained for each superoutburst are summarised in Table~\ref{tab:Etess}. 

Since there are no simultaneous multi-wavelength observations available for the superoutburst events, the spectral energy distribution (SED) during outburst is not directly constrained. Published spectroscopic temperature estimates for VW~Hyi from $\sim 20{,}000$,K to $\sim 50{,}000$,K \citep[e.g.,][]{Sion_1995, Godon_2004, Godon_2005, Long_2009}. This uncertainty has a profound impact on any attempt to infer bolometric energetics from optical photometry alone. Because the \tess\ bandpass lies on the red tail of the SED during a superoutburst, it only samples the Rayleigh–Jeans tail of the SED and as a result, the inferred bolometric correction is highly sensitive to the assumed temperature: adopting $T_{\rm eff}=20{,}000$K yields a bolometric correction of $\sim 34$, while a  $T_{\rm eff}=50{,}000$K implies $\sim320$, an order-of-magnitude difference.  We therefore restrict our analysis to band-limited energetics within the \tess\ passband. 

\subsection{Masses of the White Dwarf and Donor Stars}
\label{sec:component_masses}
We estimate the component masses of VW~Hyi by combining (i) the superhump-based mass ratio derived from the \emph{Stage~A} superhump frequency excess, which provides a dynamical
estimate of $q$, and (ii) external constraints on either the white-dwarf mass or the donor mass. As discussed in Section~\ref{sec:superhump_periods_mass_ratios}, our Stage~A analysis yields $q_A = 0.1341 \pm 0.0047$ (SO1), $0.1296 \pm 0.0054$ (SO2), and $0.1306 \pm 0.0024$ (SO3). Adopting the weighted mean of these independent measurements, we use a representative Stage~A mass ratio of $q = 0.1311 \pm 0.0020$, where the uncertainty reflects propagation of the Stage~A period uncertainties.

White dwarfs in short-period SU~UMa-type cataclysmic variables cluster around $M_1 \simeq 0.81 \pm 0.20\,M_\odot$ \citep{Zorotovic:2011, Godon_2021, Pala_2022}. Adopting this empirical prior together with our representative Stage~A mass ratio ($q = 0.1311 \pm 0.0020$) yields a donor mass
\[
M_2 = q\,M_1 = 0.106 \pm 0.026\,M_\odot,
\]
where the uncertainty reflects the quadrature combination of the dispersion in $M_1$ and the propagated uncertainty in $q$. Even adopting an extremely conservative interval $M_1 = 0.60$--$1.00\,M_\odot$ implies $M_2 \simeq 0.079$--$0.131\,M_\odot$. These values lie securely above the hydrogen-burning limit ($\sim0.072$--$0.075\,M_\odot$; \citealt{Chabrier_1997, Baraffe:2015}), demonstrating that the donor in VW~Hyi is a low-mass main-sequence (or slightly evolved) star rather than a brown dwarf.

An independent constraint is provided by the semi-empirical donor sequence for cataclysmic variables \citep{Knigge_2006, Knigge:2011}. At the orbital period of VW~Hyi, $P_{\rm orb} = 0.074271\,\mathrm{d}$ ($\simeq107\,\mathrm{min}$), this sequence predicts a donor mass $M_2 \simeq 0.12$--$0.13\,M_\odot$ and a donor radius of $\simeq0.16$--$0.18\,R_\odot$.
Combining a representative value $M_2 = 0.125 \pm 0.015\,M_\odot$ with our Stage~A mass ratio yields a white-dwarf mass
\[
M_1 = \frac{M_2}{q} = 0.95 \pm 0.12\,M_\odot,
\]
where the quoted uncertainty incorporates both the intrinsic scatter of the donor sequence and the uncertainty in $q$. This value is consistent with the upper end of the white-dwarf mass distribution in SU~UMa systems and with ultraviolet spectral modeling of VW~Hyi reported in the literature \citep[e.g.,][]{Godon:2004}.

Taken together, both approaches converge on component masses of $M_1 \simeq 0.6$--$1.0\,M_\odot$ and $M_2 \simeq 0.08$--$0.14\,M_\odot$. The dominant sources of systematic uncertainty are (i) the external prior on the white-dwarf mass and (ii) the intrinsic dispersion of the donor sequence; the contribution from the superhump-derived mass
ratio is sub-dominant because the Stage~A superhump periods are well constrained (though additional systematic uncertainty in the Stage~A calibration may be present at the few-$\times10^{-3}$ level).

Period-bouncer cataclysmic variables are characterized by very low mass ratios ($q \lesssim 0.07$--$0.08$) and substellar donors ($M_2 < 0.075\,M_\odot$) at orbital periods $P_{\rm orb} \gtrsim 90\,\mathrm{min}$ \citep{Patterson:2011, Knigge:2011}. The mass ratio and component masses inferred for VW~Hyi are therefore inconsistent with a brown-dwarf donor at high significance and demonstrate that VW~Hyi has not evolved beyond the orbital period minimum.

\begin{figure}[ht!]
    \centering
    \includegraphics[width=0.9\textwidth]{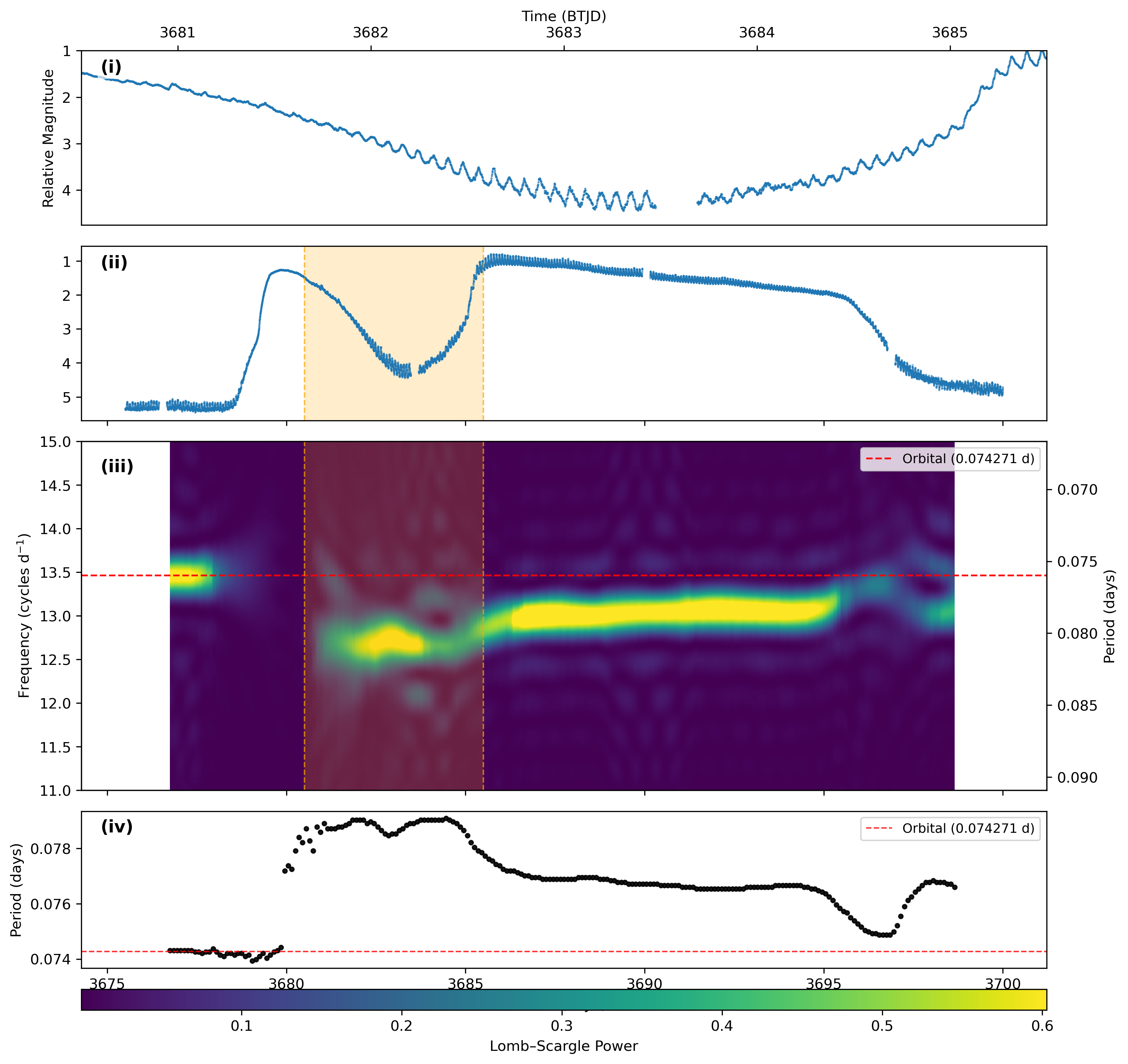}
\caption{%
    \textit{TESS} photometry of VW~Hyi during the SO2 (Sectors 87+88; time in BTJD). 
    (i) Zoomed view of the precursor--dip interval. 
    (ii) Full light curve over the analyzed time span; the shaded region marks the interval highlighted in panel (i). 
    (iii) Sliding-window Lomb--Scargle time--frequency map of the light curve in (ii), computed with a window length of 2.5\,d and a step of 0.10\,d; colors indicate Lomb--Scargle power (see color bar). The horizontal red dashed line marks the orbital signal ($P_{\rm orb}=0.074271$\,d), and the right-hand axis shows the corresponding period. 
    (iv) Period corresponding to the maximum Lomb--Scargle power in each window as a function of time, illustrating the evolution of the dominant modulation relative to the orbital period.%
}
    \label{fig:SO2_VW_Hyi_dynamic_LS}
\end{figure}

\begin{figure}[ht!]
    \centering
    \includegraphics[width=0.9\textwidth]{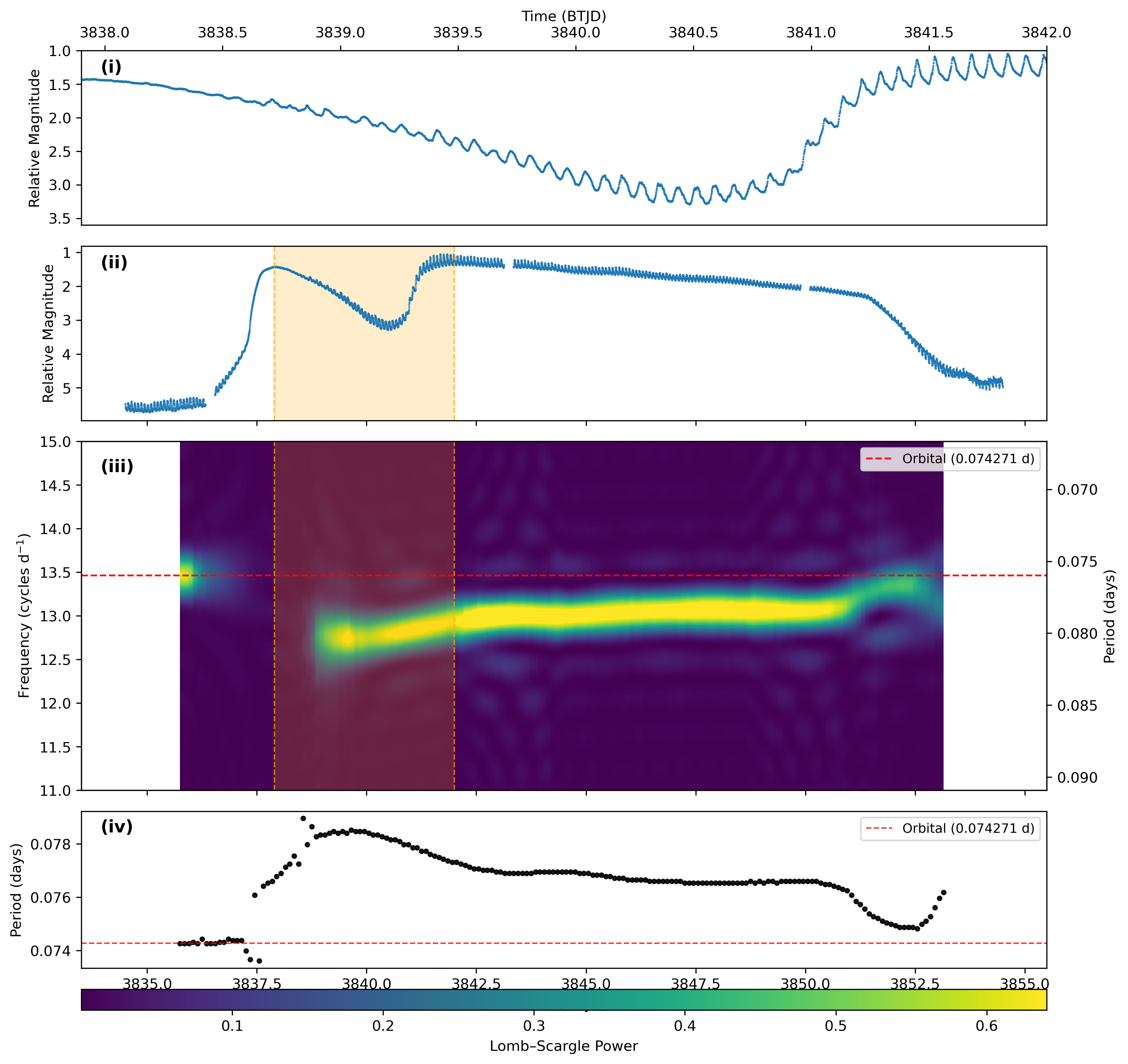}
    \caption{%
    Same as Figure~\ref{fig:SO2_VW_Hyi_dynamic_LS}, but for the SO3 observed in \textit{TESS} Sector~93. 
    Panels (i)--(iv) show, respectively, a zoomed view of the precursor--dip interval, the full light curve, the sliding-window Lomb--Scargle time--frequency map (window length 2.5\,d; step 0.10\,d), and the corresponding best-fitting period evolution. 
    The red dashed line marks the orbital period ($P_{\rm orb}=0.074271$\,d), and the shaded region highlights the interval of enhanced period variability during the decline from the precursor.%
    }
\label{fig:SO3_VW_Hyi_dynamic_LS}
\end{figure}

\section{Discussion: Comparisons and Broader Implications}
\label{sec:discussion}

The \tess\ light curves of VW~Hyi reveal a deep, temporally distinct precursor--dip that precedes the onset of the main superoutburst plateau. VW~Hyi therefore joins the small set of SU~UMa-type dwarf novae in which a well-separated precursor--dip--main sequence has been previously reported, including the \textit{Kepler} systems V344~Lyr and V1504~Cyg \citep{Still_2010, Osaki_Kato_2013, Osaki_Kato_2014} and the ground-based cases QZ~Vir \citep{Imada_2017} and PM~J03338+3320 \citep{Kato_2016}. For the three superoutbursts analyzed here, the precursor morphology varies from event to event, with SO2 showing the deepest separated dip. In SO2 and SO3, where the precursor dip is well separated, the superhump growth begins during the fading phase and continues through the rise into the main plateau. This overall sequence is naturally understood within the \tti framework \citep[e.g.,][]{Osaki_Kato_2013, Osaki_Kato_2014}. The uninterrupted \tess\ coverage further enables a detailed comparison to the \textit{Kepler} phenomenology, revealing both shared characteristics and key differences that broaden the observed diversity of precursor events.

\subsection{An unusually deep and long-lived precursor dip}
The precursor dip in VW~Hyi  is notably deeper and longer-lived than those reported in V344~Lyr and V1504~Cyg, most prominently during the Sector~87 event (SO2). In the \tess\ data, the flux during the dip approaches the quiescent level and remains suppressed for $\sim$4 (in SO2) and $\sim$ 2~days (in SO3). The full precursor-to-plateau evolution is correspondingly extended, with peak-to-peak durations of $\sim$6~days (SO2) and $\sim$4~days (SO3). The rise from the dip minimum to the plateau is also markedly steep; for SO2 we measure a rise rate of $\sim-2.35\pm0.02$~mag~day$^{-1}$ over the highlighted interval.

Within the \tti\ picture, a deep, extended dip can occur if the disk reaches the hot (outburst) state and expands toward the 3:1 resonance, but the tidal response and eccentricity growth develop only slowly. In such a case, a cooling wave can propagate substantially and drive the system close to quiescence before strengthening tidal torques halt the cooling and re-establish a hot outer disk; the subsequent steep rise reflects the rapid onset of efficient tidal dissipation once a sufficiently large fraction of the outer disk participates in the resonance \citep[e.g.,][]{Osaki_Kato_2013, Osaki_Kato_2014}. Recent time-dependent thermal--tidal simulations by \citet{Jordan_2024} qualitatively support intermediate phases in which luminosity declines temporarily before tidal amplification drives the system into a sustained superoutburst state and superhumps emerge. In this sense, the deep precursor-dip occupies an intermediate regime between (i) shallow precursor dips, where tidal dissipation suppresses cooling early, and (ii) ``failed'' precursor-like episodes, in which cooling succeeds before eccentricity growth becomes established as found in V1504~Cyg \citep{Osaki_Kato_2014}.

\subsection{Superhump growth through the dip and into the plateau}
A key commonality with the \textit{Kepler} is that superhump power in VW~Hyi becomes evident during the declining branch of the precursor (Panel (i) of Figures ~\ref{fig:SO2_VW_Hyi_dynamic_LS} and \ref{fig:SO3_VW_Hyi_dynamic_LS}). Our time--frequency analysis further shows that the dominant period (positive superhumps) evolves smoothly across the dip and into the plateau: significant periodic power is already present during the decline, the period increases gradually across the dip, reaches the maximum, the decreases slightly again before it stabilizes only after the system re-establishes the hot state, coincident with a rapid increase in signal coherence and amplitude (panels (iii) and (iv)  of  Figures ~\ref{fig:SO2_VW_Hyi_dynamic_LS} and \ref{fig:SO3_VW_Hyi_dynamic_LS}).

We interpret this behavior as evidence that disk eccentricity begins to grow during the declining branch of the precursor, while a stable Stage~A superhump---associated with coherent dynamical precession near the 3:1 resonance---is established only after the system returns firmly to the hot state. This sequence is fully consistent with the \tti\ model and closely resembles the phenomenology emphasized by \citet{Osaki_Kato_2013} for deep-precursor events. For comparison, in V1504~Cyg the rise into the superoutburst can be more discontinuous, which has been interpreted as reflecting difficulty in re-establishing the hot state when tidal dissipation is initially weak.

\subsection{Energetics}
Despite the differences in precursor depth and plateau duration in the analysed superoutbursts, our \tess-band fluence estimates (Section~\ref{sec:energetics}) indicate that the band-limited radiated energies are comparable to within uncertainties across the three events. If the \tess-band energy output is approximately conserved while durations vary, then the mean \tess-band luminosity must adjust accordingly, implying that longer plateaus are, on average, dimmer in this band. Within DIM/\tti\ language, this behavior can be accommodated if the amount of mass participating in strong tidal dissipation at any given time differs between events, and/or if tidal dissipation turns on later or operates less efficiently following the precursor. Because the bolometric correction during outburst is uncertain and the \tess\ band samples the red tail of the outburst SED, these inferences should be understood primarily as constraints on the \emph{optical} (band-limited) energetics rather than definitive bolometric energy budgets.

\subsection{Are We Observing the Thermal--Tidal Instability?}
\label{sec:tti_test}

A central question raised by deep precursor--dip superoutbursts is whether the dip marks the onset and growth of the tidal instability envisioned in the \tti model. In this framework, the precursor outburst drives the accretion disk to a larger radius; once the outer disk reaches the 3:1 resonance, enhanced tidal torques excite eccentricity and produce superhumps, while the accompanying increase in tidal dissipation helps sustain the hot disk into the main superoutburst plateau \citep[e.g.,][]{Osaki_1989, Osaki_Kato_2013, Osaki_Kato_2014}. In this picture, a pronounced dip can occur if a cooling transition begins after the precursor but is rekindled as the tidal instability strengthens and the disk is re-heated into the plateau.

The \tess\ observations of VW~Hyi provide a time-resolved test of this sequence. In the SO2 and SO3 events, significant power near the superhump frequency is already detectable during the decline into the precursor minimum and then amplifies as the system rises into the main plateau. At the same time, the signal becomes progressively more coherent once the hot state is re-established, consistent with a transition from weak or spatially intermittent eccentricity to a stable Stage~A superhump pattern associated with coherent dynamical precession near the 3:1 resonance. The key discriminant is thus not the dip morphology alone, but the observed timing: superhump power emerges during the dip recovery and strengthens continuously into the plateau, as expected if the disk reaches the resonance during the precursor and the tidal instability develops while the thermal state recovers.

The evolution of the superhump periods through the deep further supports this interpretation. Using the Stage~A superhump periods measured during the early plateau, we infer a mass ratio of $q\simeq0.12$--0.13 (Section~\ref{sec:superhump_periods_mass_ratios}), consistent with independent estimates for VW~Hyi and firmly within the low-$q$ regime in which the 3:1 resonance is accessible. Taken together, the emergence of superhump power during the precursor decline, its smooth amplification into the plateau, and the low inferred $q$ strongly favor a tidal origin for the dip and subsequent superoutburst evolution.

While precursor--dip morphologies can, in principle, be produced in models beyond \tti, the key issue is whether those models also reproduce the \emph{phase-resolved emergence} of the superhump signal during the declining phase of the dip. In the enhanced mass-transfer (EMT) picture \citep[e.g.,][]{Smak_2004, Smak_2008}, a long plateau is maintained primarily by a transient increase in $\dot{M}$ from the donor, whereas in ``pure'' thermal limit-cycle interpretations \citep[][]{Cannizzo_2010, Cannizzo_2012}, early structure can arise from heating-front propagation and temporary stalling within the disk. More recently, disk-instability models augmented with magnetically driven wind torques have been shown to generate dip-like features by enhancing angular-momentum loss and partially depleting the disk \citep{Scepi_2019}. A generic limitation of these alternatives, however, is that the dip is not intrinsically tied to reaching the 3:1 resonance, and thus they do not naturally predict a tight temporal connection between the dip and the onset of a superhump. Any EMT- or wind-torque-based explanation for VW~Hyi must therefore account not only for the light-curve morphology, but also for the observed amplification and increasing coherence of superhump power through the dip and into the plateau. Moreover, there is no independent evidence that VW~Hyi hosts a strongly magnetic primary that would be expected to imprint clear spin modulations or polarimetric signatures \citep[e.g.,][]{Ferrario_1993, Woudt_2012}. We therefore interpret the deep precursor--dip in VW~Hyi as the natural outcome of a precursor outburst followed by the onset and growth of the tidal instability, and identify the dip-phase growth of the superhump signal as the most direct observational discriminator for confronting time-dependent \tti\ simulations with precision \tess\ photometry. A detailed model-by-model discussion of these competing interpretations in the context of \textit{Kepler} dwarf novae is given by \citet{Osaki_Kato_2013}.

\section{Conclusion}
\label{sec:conclusion}

We have presented high-cadence \tess\ observations of three well-resolved superoutbursts of the SU~UMa-type dwarf nova VW~Hyi, two of which (SO2 in Sectors~87+88 and SO3 in Sector~93) exhibit a pronounced, temporally distinct precursor--dip--main plateau morphology. This behavior, predicted by the  \tti model, has previously been documented in \textit{Kepler} observations of V1504~Cyg and V344~Lyr. Continuous space-based coverage at 120~s cadence enables us to resolve the superoutburst sequence in VW~Hyi with unprecedented detail,including a deep dip approaching quiescence in SO2 and SO3, followed by a rapid rise into the main plateau.

We measure precise timings between the precursor peak, dip minimum, and plateau onset that are comparable to those in the \textit{Kepler} archetypes. A time--frequency analysis reveals significant periodic power near the superhump frequency already during the late precursor decline, followed by a smooth evolution and stabilization of the superhump period after the dip. This dip-phase emergence and subsequent increase in coherence are consistent with the growth and saturation of disk eccentricity at the 3:1 resonance envisioned in the \tti\ framework.

Using the stabilized Stage~A superhump periods and the frequency-excess calibration of the dynamical precession rate, we infer a mass ratio $q \simeq 0.12$--0.13 for VW~Hyi. This value is consistent with, but more robustly determined than, previous estimates based on ground-based data and empirical Stage~B relations. The inferred mass ratio implies a main-sequence donor with $M_2 \gtrsim 0.12\,M_\odot$ and a white dwarf of mass $\sim0.6$--$0.8\,M_\odot$, ruling out a brown-dwarf donor and confirming that VW~Hyi has not evolved beyond the orbital-period minimum. Our result agrees well with previous Stage~A determinations \citep[e.g.,][]{Kato_2013a, Kato_2022}, supporting the robustness of the dynamical-precession method and reinforcing VW~Hyi as a low-$q$, non--period-bouncer SU~UMa system.

Despite notable event-to-event differences in precursor--dip depth and superoutburst duration, the band-limited radiated energies measured in the \tess\ passband are consistent within uncertainties, indicating a tight coupling in the optical output across events. Within the \tti\ picture, this is naturally understood if the disk reaches---and only marginally exceeds---the 3:1 resonance during the precursor, triggering the tidal instability, while the mass reservoir available for accretion and tidal dissipation during the plateau remains broadly similar from cycle to cycle.

Deep, well-resolved precursor dips remain rare, likely because their secure identification requires high precision and near-continuous coverage across the onset of a superoutburst---conditions most readily met by \textit{Kepler} and \tess. They may also be physically favored in low-$q$ systems, where disks can store substantial mass at large radii, allowing the disk to approach the resonance while the growth of eccentricity and associated tidal dissipation are temporarily delayed. VW~Hyi satisfies these conditions, making it an especially favorable laboratory for probing the onset of the tidal instability.

Taken together, the \tess\ observations of VW~Hyi provide one of the clearest views to date of the transition from a precursor outburst to a tidally driven superoutburst and establish VW~Hyi as a benchmark system for confronting time-dependent \tti\ simulations with precision photometric constraints. Future coordinated multi-wavelength campaigns---combining \tess-like optical coverage with ultraviolet observations such as the upcoming ULTRASAT mission \citep{Shvartzvald_2024}, and X-ray monitoring---will enable the inner and outer disks to be tracked simultaneously and offer deeper insight into the physical processes governing superoutbursts in low-$q$ cataclysmic variables.

\section*{Acknowledgments}
This work was supported in part by the Israel Ministry of Innovation, Science, and Technology under Grant No.~0008107 and by the Israel Science Foundation (ISF) under Grant No.~1404/22. This paper includes data collected by the \tess\ mission, which are publicly available from the Mikulski Archive for Space Telescopes (MAST). Funding for the \tess\ mission is provided by NASA’s Explorer Program. The Space Telescope Science Institute (STScI) is operated by the Association of Universities for Research in Astronomy, Inc., under NASA contract NAS~5--26555.
B.B.O. acknowledges the hospitality of the Kavli Institute for Astrophysics and Space Research and the \tess\ Science Office at the Massachusetts Institute of Technology during a portion of this work.

This research made use of \texttt{Astropy} \citep{astropy:2013, astropy:2018, astropy:2022}, a community-developed core Python package for astronomy. All figures were generated using \texttt{Matplotlib} \citep{Hunter:2007}.

\bibliography{references}{}
\bibliographystyle{aasjournal}

\end{document}